\documentclass[useAMS,usenatbib]{mnras} 
\usepackage{graphicx}
\usepackage[caption=false]{subfig}
\usepackage{amssymb, amsmath}
\usepackage{times}
\usepackage{color}
\usepackage{lscape}
\usepackage[section]{placeins}
\usepackage{breqn}
\usepackage{hyperref}

\usepackage{lineno}

\usepackage{arydshln}

\usepackage{url}

\newcommand{\be}{\begin{equation}}
\newcommand{\ee}{\end{equation}}

\newcommand{\h}{\mathcal{H}}

\newcommand{\gtsima}{$\; \buildrel > \over \sim \;$}
\newcommand{\ltsima}{$\; \buildrel < \over \sim \;$}
\newcommand{\simgt}{\lower.7ex\hbox{\gtsima}}
\newcommand{\simlt}{\lower.7ex\hbox{\ltsima}}
\newcommand{\zshell}{z_{\rm g}^{\rm shell}}

\begin{document}


\title{Relativistic asymmetries in the galaxy cross-correlation function}

\author[Giusarma et al.]{Elena Giusarma$^{1,2,3,4}$
\thanks{E-mail: \href{mailto:mn@ras.org.uk}{egiusarm@andrew.cmu.edu}}
, Shadab Alam$^{1,2, 5}$
\thanks{\href{mailto:mn@ras.org.uk}{salam@roe.ac.uk }}
,  Hongyu Zhu $^{1,2}$, Rupert A. C. Croft$^{1,2}$, \newauthor and Shirley Ho$^{1,2,3,4}$\\
   $^{1}$ Department of Physics, Carnegie Mellon University, 5000 Forbes Ave., Pittsburgh, PA 15213, USA \\
    $^{2}$ McWilliams Center for Cosmology, Carnegie Mellon University, 5000 Forbes Ave., Pittsburgh, PA 15213, USA \\
    $^{3}$ Berkeley Center for Cosmological Physics, University of California, Berkeley, CA 94720, USA \\
    $^{4}$ Lawrence Berkeley National Laboratory (LBNL), Physics Division, Berkeley, CA 94720, USA \\
    $^{5}$ Institute for Astronomy, University of Edinburgh, Royal Observatory, Blackford Hill, Edinburgh, EH9 3HJ, UK
}
\date{\today}
\pagerange{\pageref{firstpage}--\pageref{lastpage}}   \pubyear{2015}
\maketitle
\label{firstpage}

\begin{abstract}
  We study the asymmetry in the two-point cross-correlation function
  of two populations of galaxies focusing in particular on the
  relativistic effects that include the gravitational redshift. We
  derive the cross-correlation function on small and large scales
  using two different approaches: General Relativistic and Newtonian
  perturbation theory. Following recent work by Bonvin et al.,
  Gazta\~{n}aga et al. and Croft, we calculate the dipole and the
  shell estimator with the two procedures and we compare our results.
  We find that while General Relativistic Perturbation Theory (GRPT)
  is able to make predictions of relativistic effects on very large,
  obviously linear scales ($r>50$ Mpc/h), the presence of
  non-linearities physically occurring on much smaller scales (down to
  those describing galactic potential wells) can strongly affect the
  asymmetry estimators. These can lead to cancellations of the
  relativistic terms, and sign changes in the estimators on scales up
  to $r \sim 50$ Mpc/h.  On the other hand, with an appropriate
  non-linear gravitational potential, the results obtained using
  Newtonian theory can successfully describe the asymmetry on smaller,
  non-linear scales ($r<20$ Mpc/h) where gravitational redshift is the
  dominant term.  On larger scales the asymmetry is much smaller in
  magnitude, and measurement is not within reach of current
  observations. This is in agreement with the observational results
  obtained by Gazt\~{n}aga et al. and the first detection of
  relativistic effects (on ($r<20$ Mpc/h) scales) by Alam et al.
  \textcolor{red}{}
\end{abstract}


\section{Introduction}

The galaxy two-point correlation function is the most common
statistical method used to study large-scale structure.
In standard $\Lambda$CDM the Universe is homogeneous and
isotropic. This implies that the correlation function of a galaxy
population is symmetric under the exchange of two galaxies. However,
the observed correlation function in galaxy redshift survey is not
isotropic. The anisotropy arises because the distance is estimated
from redshift which is affected by a number of physical processes.
The most prominent effect is due to the peculiar velocities of galaxies, known as redshift space distortion (RSD)~\citep{davis1983, Kaiser01071987, Hamilton:1997zq}, which produces a quadrupole moment of the correlation function, but not a dipole. Other relativistic effects, including gravitational potential, can induce higher order subtle anisotropy and asymmetry in the correlation function. 
The monopole and quadrupole have been measured in  different galaxy surveys~\citep{Alam2015, Beutler:2016ixs}, and they provide a tool to test cosmological models, constrain cosmological
parameters and derive measurements of the growth rate of structure~\citep{blake2013, samushia2014, Alam2015, Alam2016Testing}.

If one considers galaxies in two different populations, for example with different luminosities or masses,  the correlation function also assumes
 an anti-symmetric part. The breaking of symmetry generates an odd multipole, namely dipole. 
The asymmetry in the two-point correlation function measured from large-scale structure has been discussed by many authors including:~\cite{McDonald:2009ud},~\cite{yoo2009},~\cite{yoo2012} ,~\cite{Yoo:2014sfa},~\cite{Croft2013},~\cite{Bonvin:2014owa},~\cite{Bonvin2014b} and~\cite{Zhu2016Nbody}.
\raggedbottom

There are several physical processes associated with galaxy properties, dynamics and environment which can affect the observed redshift and lead to a non-zero dipole moment in the cross-correlation function.~\cite{Bonvin:2014owa} and~\cite{Bonvin2014b} discuss three of these effects that generate the asymmetry in
the cross-correlation function. The first one is the~\textit{relativistic effect} which includes, among others, the gravitational redshift and 
Doppler term. The gravitational redshift occurs when a photon loses its energy escaping from a potential well and this leads to an increase in its wavelength. fThis distortion is therefore 
induced by a gradient of the gravitational potential.  
The Doppler term depends on the relative motion of the galaxy with respect to the observer: the apparent magnitude and luminosity of the galaxy change, depending on whether it appears closer or farther away
from the observer in redshift space. 
The second contribution is due to the~\textit{evolutionary effect} and it arises from the redshift evolution terms in the bias and the growth function.
The last contribution generating an asymmetry in the two-point correlation function is the~\textit{wide-angle effect} due the fact 
that we are observing on a cone and therefore there is a difference in the line-of-sight directions between any two galaxies.
In the distant-observer approximation the separation angle between two galaxies is small and this makes the two line-of-sight directions
to the galaxies parallel. The wide-angle effect results when this approximation
is no longer accurate.

In this paper we provide theoretical predictions for the asymmetry in the two-point cross-correlation function focusing in particular on the gravitational redshift effect.

The gravitational redshift was predicted by Einstein during the
development of the General Theory of Relativity
\citep[GR]{Einstein1916} and was measured for the first time
by~\cite{Pound1959}.  Subsequently other determinations have been
realized in astrophysical
environments and the Solar systems~\citep{Greenstein1971, Lopresto1991} and in galaxy clusters~\citep{Wojtak2011}.\\
\cite{Cappi1995} studied the gravitational redshifts in galaxy
clusters assuming different density profiles and showing that a non
negligible effect can be predicted in individual rich
clusters.~\cite{Kim2004} proposed to use surveys of cluster galaxies
to obtain a statistical measurement of the gravitational redshift
profile. On larger scales, ~\cite{McDonald:2009ud} explored the effect
of gravitational redshifts on the cross-power spectrum of two
different populations of galaxies using linear perturbation
theory.~\cite{Croft2013} used a halo model to predict the relative
gravitational redshifts between simulated galaxies including
non-linear scales.  Good agreement was found between the halo model
and N-body simulations and an estimator was introduced to quantify the
line-of-sight asymmetry in the cross-correlation function of the two
galaxy samples.~\cite{Bonvin2014b} made use of GR perturbation theory
to study the different contributions that induce the asymmetry in the
two-point cross-correlation function of two populations of galaxies
focusing in particular on the relativistic contribution and
considering large-scale structure on linear
scales.~\cite{Gaztanaga:2015} measured the distortions in the
cross-correlation function from the Baryon Oscillation Spectroscopic
Survey (BOSS) Data Release 10 CMASS sample looking on large, linear
scales ($r>20$ Mpc/h), finding consistent results with theory,
but no evidence for a signal. 
~\cite{Alam2016Measurement} presented a first
detection of the  redshift asymmetry at 2.7 $\sigma$ in the
cross-correlation function using the Sloan Digital Sky Survey (SDSS)
Data Release 12 (DR12) CMASS sample of galaxy. The signal peaks on
small scales ($r<10$ Mpc/h) where the linear perturbation theory starts to
be less accurate.~\cite{Zhu2016Nbody} used a quasi-Newtonian approach
with N-body simulations in order to explore the relativistic
distortions of galaxy clustering focusing on non-linear scales ($r<20$
Mpc/h).~\cite{Alam2016TS} looked at analyzed relativistic beaming
effect in SDSS CMASS DR12.

In this work we study the cross-correlation asymmetry of two galaxy
populations using two different approaches: General Relativistic
Perturbation Theory (GRPT), which robustly describes the relativistic
clustering on large, linear scales, and Newtonian perturbation theory,
which is used to model relativistic effects on small non-linear
scales. For the latter, we limit our modeling to the gravitational
redshift, and the reader is encouraged to consult ~\cite{Zhu2016Nbody} for
a full treatment which includes the other Doppler and higher order
terms in a simulation context. 
Our aim in this work is to compare the GRPT and Newtonian approaches in order
to obtain a theoretical model applicable to all scales.

This paper is organized as follows. Based on recent works of~\cite{Bonvin2014b} and~\cite{Gaztanaga:2015}, in Section~\ref{sec:theory} we describe the formalism used
to derive the dipole in the two-point correlation function and we present the shell estimator 
introduced by~\cite{Croft2013}. In Section~\ref{sec:computing} we compute the dipole and the shell
estimator for the relativistic and wide-angle contributions considering different clustering biases for the galaxy populations and using GRPT.
In Section~\ref{sec:distortion} we study the distortion of the galaxy cross-correlation function induced
by gravitational redshift and peculiar velocities on all scales making use of Newtonian perturbation theory and we compare our results using the two approaches. Finally, we discuss and draw our
conclusions in
Section~\ref{sec:discussion}.


  \section{General Relativistic perturbation theory approach}
\label{sec:theory}

To extract the anti-symmetric term in the two-point correlation
function we follow the approach adopted by~\cite{Bonvin:2015}
and~\cite{Gaztanaga:2015} to split the galaxies into two populations
according to their luminosity and mass, separated by a distance
$r$. We indicate as $g1$ the bright (B) population with higher mass
and as $g2$ the faint (F) population with lower mass.

The anti-symmetric part of the cross-correlation function between these two populations is given by: 
\begin{equation}
\xi_{g1g2}(z,z')=\langle\delta_{g1}(\mathbf{x},z)\delta_{g2}(\mathbf{x}',z')\rangle~\neq\xi_{g2g1}(z',z)~,
\end{equation}
in which $z$ ($z'$) and $\mathbf{x}$ ($\mathbf{x}'$) are the redshift and the position of the bright (faint) galaxy, $\delta_{g1}(\mathbf{x},z)$ 
and $\delta_{g2}(\mathbf{x}',z')$ are the respective over-densities. The term $\xi_{g1g2}(z,z')$ 
is the sum of three different contributions: the \textit{relativistic distortion}, the \textit{evolution} of the bias and growth rate, and the \textit{wide-angle} effect. If we expand this term in odd multipoles of the cosine of the angle that the pair makes with the observer's line-of-sight \citep{Bonvin2014b, Bonvin:2014owa, Raccanelli:2013dza}, $\cos\alpha_{i,j}$, we can note that the dominant contribution to the distortion is due to the dipole. 
As shown in~\cite{Gaztanaga:2015} and~\cite{Bonvin:2015} the derivation of the dipole depends on the choice of kernel $W_{\mathbf{x}_i,\mathbf{x}_j,L_i, L_j}$  which must be anti-symmetric under the
exchange of $i$ and $j$, where $i$ and $j$ are the cells in which the survey is pixelized and in which the galaxy over-densities are defined. 
Moreover, it must depend on the luminosity of each pixel, $L_i$ and $L_j$, and  on $\cos\alpha_{i,j}$ . Therefore we can write the general expression for the dipole as follows, (see~\cite{Gaztanaga:2015} for for more details): 
\begin{equation}
\hat{\xi}=\sum_{i,j}\sum_{\rm{B,F}}W_{\mathbf{x}_i,\mathbf{x}_j,\rm{B, F}}\delta n_{\rm{B}}(\mathbf{x}_i)\delta n_{\rm{F}}(\mathbf{x}_j)~,
\end{equation}
in which we substitute $L_i,L_j=\rm{B,F}$, while $\delta n_{\rm{B}}(\mathbf{x}_i)$ and $\delta n_{\rm{F}}(\mathbf{x}_j)$ denote the over-density of galaxies for each pixel. 

We now define the dipole for the different contributions that generate the anti-symmetry in the  two-point correlation function.

\subsection{Contributions to the dipole}
In the continuous limit, the dipole contribution due to the relativistic effect is given by (see~\cite{Bonvin2014b} for a complete derivation): 
\begin{equation}
\begin{split}
\langle\hat\xi_{\rm rel}(r)\rangle&=(b_B-b_F)\frac{f}{2\pi^2}\left(\frac{\dot{{\mathcal{H}}}}{{\mathcal{H}}^2}+\frac{2}{\chi{\mathcal{H}}}\right)\frac{\h}{\h_0}\\
&\times\int dk k \h_0 P(k,\bar{z})j_1(kr)~,\label{rel}\\
\end{split}
\end{equation} 
in which $\chi$ is the comoving distance, $b_B$ and $b_F$ are the bias of bright and faint population of galaxies, $f\equiv d \ln D/d \ln a$ is the logarithmic derivative of the linear growth factor, $D$, with respect to the scale factor, $a$, $P(k,\bar{z})$ is the linear matter power spectrum at the mean redshift of the survey and $ j_1(kr)$ is the spherical Bessel function for $\ell=1$. 
We can see that the relativistic contribution depends on the bias difference $b_B-b_F$ and it vanishes when $b_B=b_F$. This means that if the bias of bright galaxies is larger than the bias of faint galaxies, the dipole is always positive. 

The dipole due to the evolution effect can be written as:
\begin{equation}
\begin{split}
\langle\hat\xi_{\rm evol}(r)\rangle&=\frac{r}{6}\left[ (b_B-b_F)f'-f(b'_B-b'_F)\right] \left(\nu_0(r)-\frac{4}{5} \nu_2(r)\right) \\
&+\frac{r}{2}(b_Bb'_F-b'_Bb_F)\nu_0(r)\ ,\label{evol}
\end{split}
\end{equation} 
where $\nu_\ell(r)$ are defined as: 
\begin{equation}
\nu_\ell(r)=\frac{1}{2\pi^2}\int dk k^2 P(k, \bar z) j_\ell(kr)\, ,\quad \ell= 0, 2~,\label{cl}
\end{equation}
and $j_\ell(kr)$ are the spherical Bessel functions. Notice that the evolution term depends on the evolution of the bright and faint galaxies (the prime in eq.~\ref{evol} denotes a derivative with respect to $\chi$). However, as shown in Figure 11 of~\cite{Bonvin2014b}, its contribution is negligible compared to the other terms, in particular to the wide-angle effect. For this reason we decide to neglect this effect in our analysis. 

\begin{figure}
\vspace{-0.1cm}
\centering
\includegraphics[width=7cm]{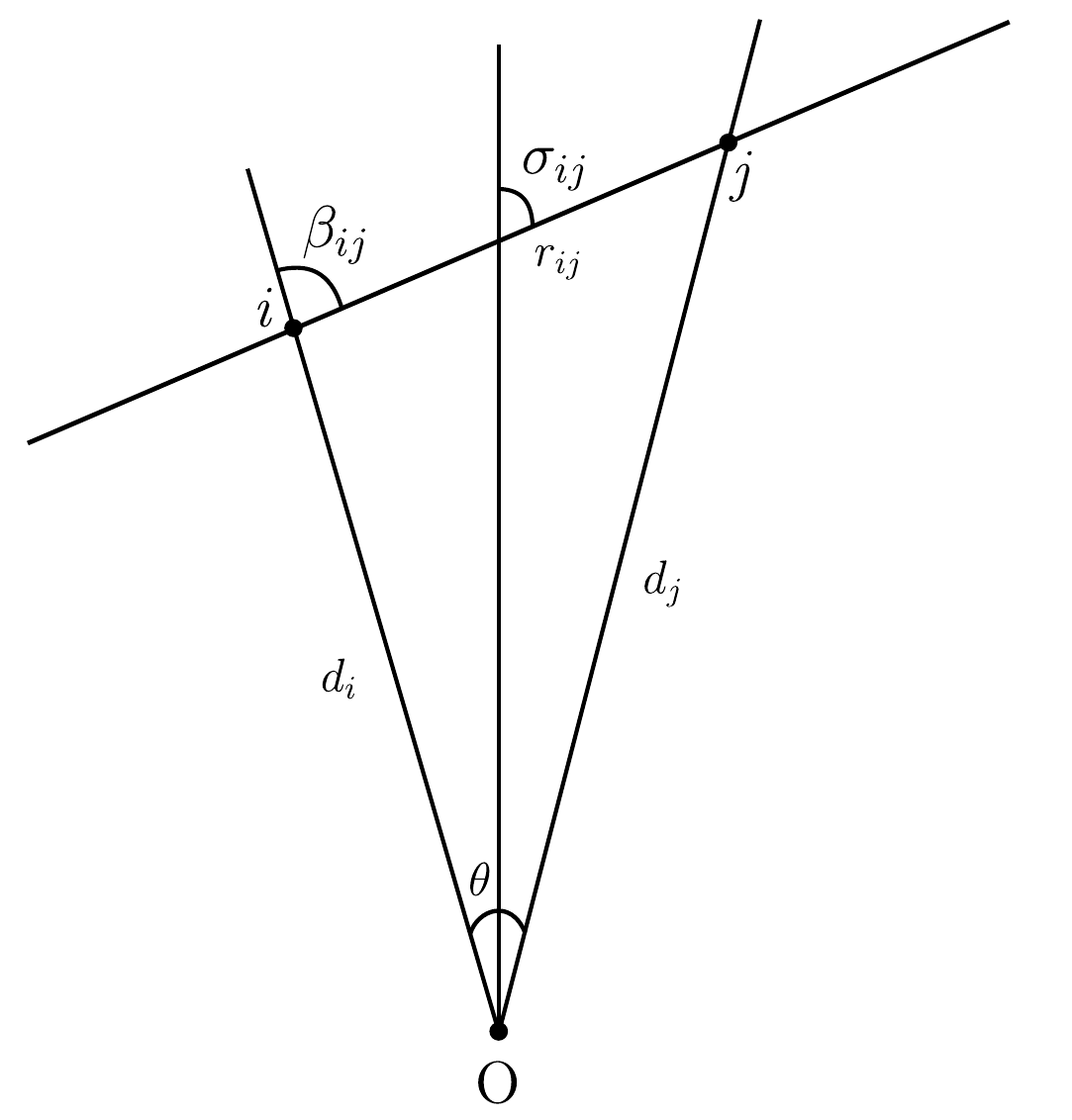} 
\caption{Coordinate system for the choice of the angle between the pair of pixels and the line-of-sight of the observer.}
\label{fig:wide1}
\end{figure}

The wide-angle effect depends on the choice of angle that the pair of pixels makes with the line-of-sight \citep{Reimberg:2015jma, Raccanelli:2010hk, Hamilton:1997zq}. We can analyze the problem within a plane formed by the two pair of pixels, $i$ and $j$, and the observer $O$, see Figure~\ref{fig:wide1}. We indicate as $d_i$ and $d_j$ the distance of pixel $i$ and pixel $j$ from the observer and with $r_{ij}$ the separation between the pair of pixels. In this plane we can consider two different choices for the angle $\alpha_{ij}$: the angle between the median and $r_{ij}$, denoted as $\sigma_{ij}$, and the angle between the direction of pixel $i$ and $r_{ij}$, denoted as $\beta_{ij}$. 
Using the $\sigma_{ij}$ angle, the expression for the dipole due to the wide-angle is: 
\be
\langle\hat\xi^{\sigma}_{\rm wide}(r)\rangle= -\frac{2f}{5}(b_B-b_F)\frac{r}{\chi}\nu_2(r)~. \label{wide},
\ee
while if we use the $\beta_{ij}$ angle, the wide-angle term is given by: 
\be
\langle\hat\xi^{\beta}_{\rm wide}(r)\rangle=\langle\hat\xi^{\sigma}_{\rm wide}(r)\rangle+\langle\hat\xi_{\rm large}(r)\rangle~,
\ee
where: 
\begin{align}
\langle\hat\xi_{\rm large}(r)\rangle=&\frac{r}{\chi}\left[b_Bb_F+(b_B+b_F)\frac{f}{3}+\frac{f^2}{5}\right]\nu_0(d)\nonumber \\
&+\frac{r}{5\chi}\left[(b_B+b_F)\frac{2f}{3}+\frac{4f^2}{7}\right]\nu_2(d)~.
\end{align} 
In this work we will only consider the wide-angle term obtained using the $\sigma_{ij}$ angle. For a complete description of the large-angle effect
see~\cite{Gaztanaga:2015}.

\subsection{Shell estimator}
The shell estimator, $\zshell$,  can be used to measure the line-of-sight asymmetry in the galaxy correlation function due to gravitational redshift.
It was introduced by \cite{Croft2013} in order to study relativistic gravitational redshift in large-scale structures. The procedure consists in binning the galaxy pair separations in spherical shell bins and then calculating the mean separation, 
$r_\parallel$, weighted by the cross-correlation funcion. 
A general expression for the shell estimator is:
\be
z_{\rm g}^{\rm shell}(r)= \frac{\int_{r'}^{r'+\Delta r'} H\left[1+\xi(r_\perp,r_\parallel)\right]r_\parallel r{^2} dr}{\int_{r'}^{r'+\Delta r'} [1+\xi(r_\perp,r_\parallel)]r{^2} dr}~,~\label{zshell1}
\ee
in which $r_\parallel$ and $r_\perp$  are the parallel and perpendicular separation between galaxies, $\xi(r_\perp,r_\parallel)$ is the two point correlation function of galaxies and $H$ is the Hubble parameter. Defining $r_\parallel$ as $r\mu$, where $\mu$ is the cosine of the angle between the pair separation $r$ and the line-of-sight, and integrating over $\mu$, we can rewrite eq.~\ref{zshell1} as follow,
\be
z_{\rm g}^{\rm shell}(r)=\frac{ \int_{-1}^{1} d\mu \int_{r'}^{r'+\Delta r'} \mu H\left[1+\xi(r,\mu)\right]r{^3} dr}{\int_{-1}^{1} d\mu \int_{r'}^{r'+\Delta r'} [1+\xi(r,\mu)]r{^2} dr}~,~\label{zshell2}
\ee
Moreover, introducing the definition of monopole ${\xi}_0(r)$ and dipole $\xi_1(r)$, eq.~\ref{zshell2} becomes:
\be
z_{\rm g}^{\rm shell}(r)=\frac{1}{3}\frac{\int_{r'}^{r'+\Delta r'} H\xi_1(r) r{^3 }dr'}{\int_{r'}^{r'+\Delta r'} [1+\xi_0(r)]r{^2} dr}~.~\label{zshell3}
\ee
In the next section we will focus on the dipole and the shell estimator of two populations of galaxies on small and large scales and we compare the results of the different contributions considering different values for the bias of bright and faint population.

\section{Computing dipole and shell estimator}
\label{sec:computing}

\begin{figure*}
 \begin{minipage}[c]{8.5cm}
   \centering
   \includegraphics[width=8.5cm]{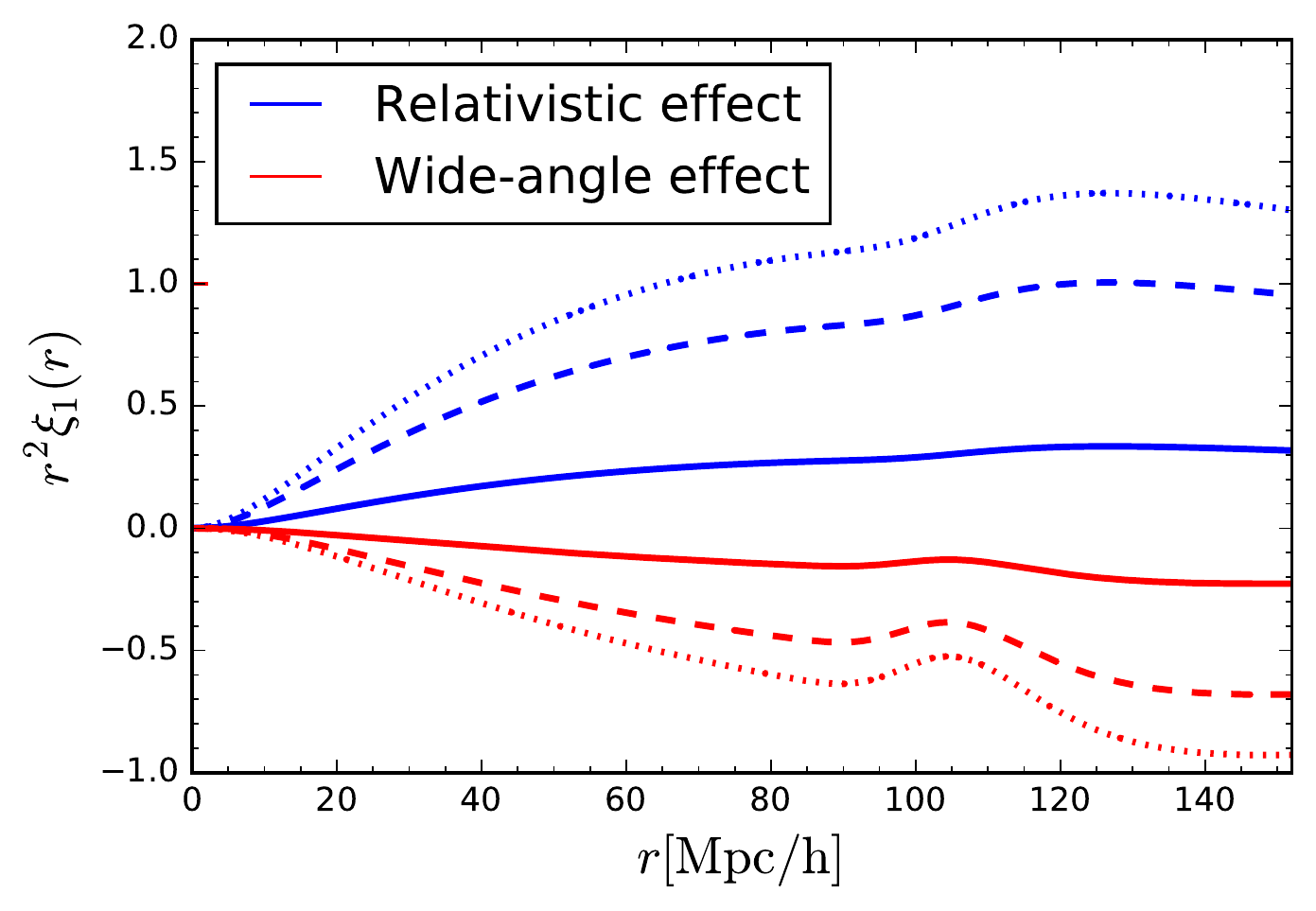}
   \caption{Dipole estimations (multiplied by $r^2$) at redshift $z=0.57$ as a function of separation $r$:
the blue and red lines refer to the relativistic and wide-angle contributions respectively.
The different linestyles are obtained assuming different values of the biases.
In particular the solid line corresponds to $b_B=2.25$ and $b_F=2.03$, the dashed line to $b_B=2.57$ and $b_F=1.91$  and the dotted line
to  $b_B=2.36$ and $b_F=1.46$.  \label{fig:dipole1}}
 \end{minipage}
 \ \vspace{2mm} \hspace{2mm} \
 \begin{minipage}[c]{8.5cm}
  \centering
   \includegraphics[width=8.7cm]{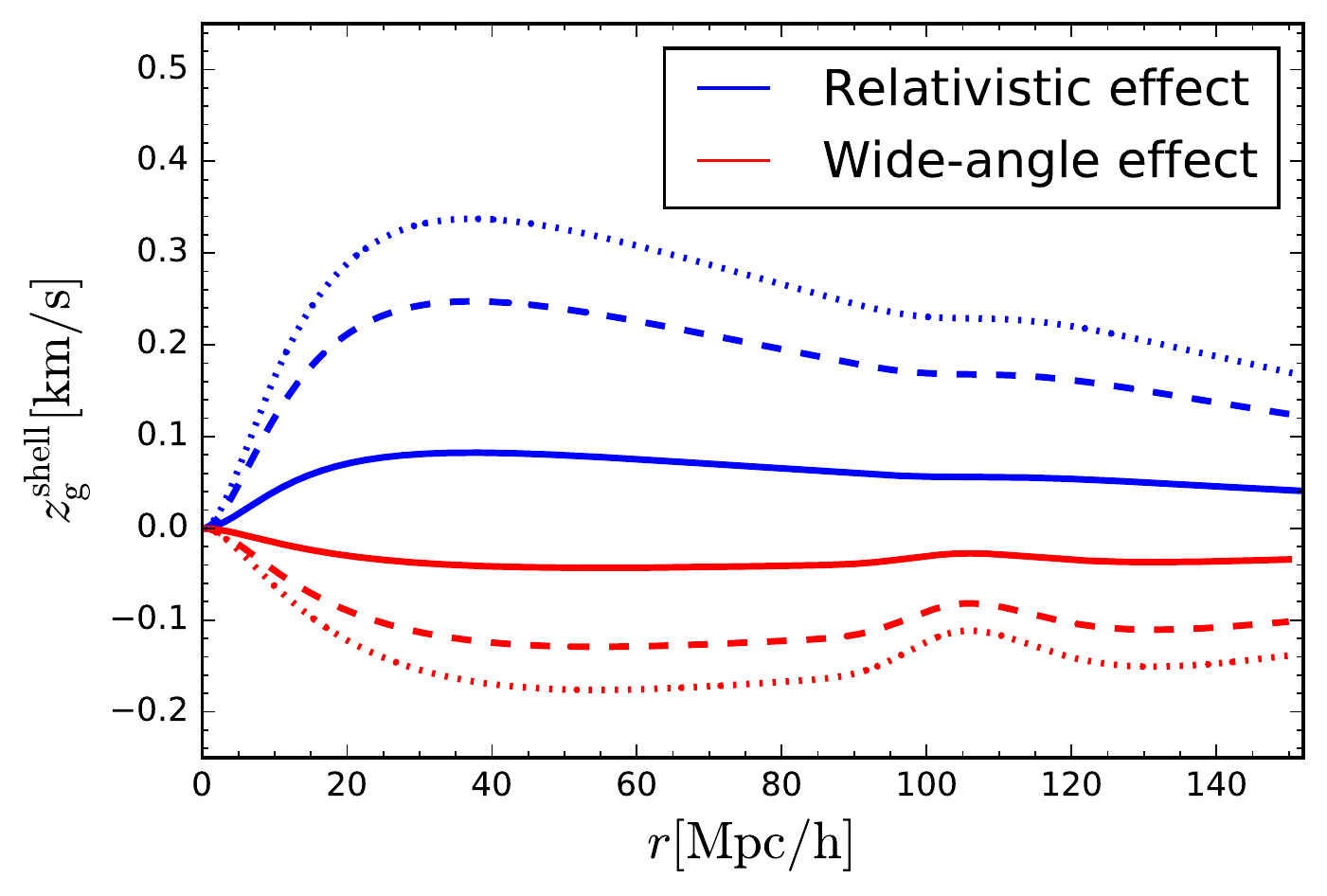}
    \caption{Shell estimations at redshift $z=0.57$ as a function of separation $r$: the blue lanes are
the relativistic contribution and the red lines are the wide-angle contribution.
Also in this case the different linestyles refer to the different values of the bias of the bright and faint populations.
See text for more details. \label{fig:zshell}}
 \end{minipage}
\end{figure*}

In this section we will compute the dipole and the shell estimator for a bright and faint population of galaxies
using the effective redshift of the 
CMASS sample of galaxies \citep{Bolton2012, Ahn2012} 
from the Baryon Oscillation Spectroscopic Survey
(BOSS) Data Release 12~\citep{Alam:2016hwk}, $z=0.57$.  
We evaluate the different contributions to the dipole and shell estimator in a flat $\Lambda$CDM model considering the 
best-fit parameters from Planck 2015 full temperature data combined with the large scale polarization measurements,~\cite{Adam:2015rua} 
($\Omega_b h^2=0.0222$, $\Omega_c h^2=0.1197$, $h=0.673$, $n_s=0.965$, $\sigma_8=0.83$).
We compute the linear matter power spectrum at redshift 0.57 using the Boltzmann code \texttt{CAMB}~\citep{Lewis:1999bs}.

In order to study the effect of the biases on the dipole and $\zshell$, we assume three different values for the bias 
of the bright and faint galaxies. We consider the biases measured by~\cite{Gaztanaga:2015} using SDSS CMASS sample Data Release 10, $b_B=2.36$ and $b_F=1.46$, and the biases calculated in~\cite{Alam2016Measurement} after dividing the SDSS CMASS DR12 sample of galaxies into two sub-samples with higher and lower mass for each photometric bands, $b_B=2.25$ and $b_F=2.03 $\footnote{We consider the two biases calculated from high-low mass subsamples cross-correlation
function in $i$ photometric band. See~\cite{Alam2016Measurement} for more details.}. Lastly we use the biases of two galaxy subsets at redshift 0.57 computed in~\cite{Zhu2016Nbody} using N-body simulations, $b_B=2.57$ and $b_F=1.91$.

Moreover in order to investigate the impact of the different contributions to the dipole and shell estimator on small and large scales, 
we consider a galaxy pair separation in the range between 0 and 150 Mpc/h. It is important to point out that a main difference between our study and the work of~\cite{Gaztanaga:2015} consists in the fact that these authors have computed the distortion in the two-point correlation function focusing on large, linear scales, i.e. $r>20$~Mpc/h.

 \subsection{Comparison of the dipoles}

In Figure~\ref{fig:dipole1} we show the different dipole contributions (relativistic and wide-angle)
as a function of the comoving separation $r$. The different linestyles of the curves correspond to
different bias values: the solid line refers to $b_B=2.25$ and $b_F=2.03$, the dashed line to $b_B=2.57$ and $b_F=1.91$   and the dotted line
to $b_B=2.36$ and $b_F=1.46$.\\
We can note that the relativistic contribution is dominant at all scales. This occurs because this term is governed by
the sign of the bias difference $b_B-b_F$ (see eq.~\ref{rel}): the bias of bright population is always
larger than the bias of the faint one and the asymmetry along the 
line-of-sight is related to the difference in the cross-correlation between the two populations of galaxies. 
As we can see from eq.~\ref{wide}, the wide-angle term also depends on the difference $b_B-b_F$ but, being negative, it is always smaller than the relativistic contribution. From these results we can deduce that 
the dipole moment could provide a powerful tool to measuring relativistic effects in galaxy clustering. 

On small scales (for $r<10$ Mpc/h), we can see that the two dipole contributions (relativistic and wide-angle) are canceled out. In particular a cancellation
of the relativistic contribution leads to no evidence for the gravitational redshift effect.
In our companion paper,~\cite{Alam2016Measurement}, we have measured the relativistic effects on the dipole moment using the cross-correlation function of two sub-samples obtained by splitting the SDSS/BOSS/CMASS sample into two equal
parts for each of the five photometric bands (\textit{u, g, r, i, z}). The results are presented in Figure 9 of~\cite{Alam2016Measurement} and they show a detection of the amplitude of relativistic asymmetry 
in each photometric band on small scales ($<$10 Mpc/h) and a zero signal on large scales ($>$10 Mpc/h). 
By comparing these with our results, we can infer that GRPT may be 
less accurate on small, non-linear scales. 
This could be due to the presence of
non-linearities that this theory is not able to treat properly.

\subsection{Comparison of shell estimator} 

Figure~\ref{fig:zshell} shows the shell estimator for the relativistic (blue lines) and wide-angle (red lines)
 contributions calculated at redshift $z=0.57$  using eq.~\ref{zshell3} and considering different values of the biases.
Note that, also for  $\zshell$, all contributions depend on the biases difference of the
two populations of galaxies and this leads the relativistic term to be dominant over the wide-angle one on all scales. 

As shown in the  figure, the wide-angle effect generates a redshift difference in $\zshell$ of -0.19 km/s for $b_B= 2.36$ and $b_F= 1.46$, of -0.12 km/sec considering
$b_B= 2.57$ and $b_F= 1.91$ and of -0.04 km/sec  using $b_B= 2.25$ and $b_F= 2.03$. If we look at very small scales ($<5$ Mpc/h),
 we can see that the wide-angle contribution is null for all the values of the biases.

On the other hand, the relativistic contribution generates a redshift difference of the order of 0.34 km/s ($b_B=2.36$ and $b_F =1.46$), 0.24 km/s ($b_B=2.57$ and $b_F =1.91$) and 0.08 km/s ($b_B=2.25$ and $b_F =2.03$) respectively. Also in this case, 
on very small scales ($r<5$ Mpc/h) the contribution of the relativistic effect to  $\zshell$ drops to 0 km/s. As pointed out by~\cite{Bonvin:2015}, a cancellation of the relativistic term means that it could be very difficult to detect the effect of gravitational redshift in the largest scale structures.

In our two companion papers~\cite{Zhu2016Nbody} and~\cite{Alam2016Measurement}, we have also computed the shell estimator for the relativistic effects of two galaxy populations. In particular, in~\cite{Zhu2016Nbody} we use N-body simulation in order to predict the gravitational redshift signal on small scales founding that the gravitational redshift effect is significant on scales of a few Mpc/h. 
In~\cite{Alam2016Measurement} we have measured the asymmetric distortions using the SDSS III  BOSS CMASS sample, and found them consistent with these
relativistic effects (where the largest contribution is from the
gravitational redshift). Specifically, we have estimated the shall estimator by fitting the theoretical model described in~\cite{Zhu2016Nbody} to the CMASS Data Release 12 using the different photometric bands. We have found a significant detection of the amplitude of gravitational redshift of 1.9$\sigma$, 2.5$\sigma$ and 1.7$\sigma$ in the $r$, $i$ and $z$ bands respectively. 
This shows that it is possible to measure the gravitational redshift in large-scale structure. 
We can conclude this section affirming that the linear perturbation theory seems to fail on small scales (where observational measurements have been made so far) because of the presence of non-linear structure formation. To account for this we need a different approach.   

\section{Newtonian perturbation theory approach}
\label{sec:distortion}
In the previous section we computed the dipole and the shell estimator generated by the relativistic and wide-angle effects
using GRPT and we showed that, on small scales, all the effects tend towards cancellation. In this section we study the effect of redshift distortions in the two-point cross-correlation function using 
Newtonian perturbation theory. In particular we examine only how the cross-correlation function of the two populations of galaxies is distorted by three
effects, gravitational redshifts, peculiar velocities or the sum of them.
As mentioned previously, we do not include all terms
that were part of the GRPT approach (such as light cone, Doppler
and wide-angle effects). As a result, we do not 
expect the Newtonian and GRPT results to be consistent on large scales, 
but we are interested in the behaviour on non-linear and quasi-linear scales.
We follow the procedure introduced by~\cite{Croft2013}.

\subsection{Defining the model}
We construct a model to describe the mean gravitational redshift difference, $\delta z_g$, between two galaxy populations, $g1$ and $g2$, and we study how this quantity, together with 
peculiar velocities, distorts the two-point correlation function. Finally we calculate the shell estimator using equation~\ref{zshell3} and we compare the results with those obtained
using GRPT.

According to General Relativity, the mean gravitational redshift difference between $g1$ and $g2$ is given by~\cite{Croft2013}:
\begin{equation}
\delta z_g= z_{g1}(0)-z_{g2}(r)= \frac{G}{c}\int_\infty^r M_{12}(x)x^{-2} dx~,
\label{zg}
\end{equation}
in which 
\begin{equation}
M_{12}(r)=4\pi\bar{\rho}\int_0^r(\xi_{g1\rho}(x)-\xi_{g2\rho}(x))x^{2} dx~,
~\label{mass}
\end{equation}
is the difference in mass, $\bar{\rho}$ is the mean density of the Universe, $G$ is the Newton's gravitational constant and $c$ is the speed of light. In eq.~\ref{mass},  
$\xi_{g1\rho}$ and $\xi_{g2\rho}$ are the $g1$ and $g2$ galaxy-mass cross-correlation functions defined as: 
\begin{equation}
\xi_{g\rho}(r)=b\xi( r)~,
\end{equation}
where $b$ is the bright or faint bias of the two populations of galaxies 
and $\xi(r)$ is the linear $\Lambda$CDM correlation function.

\begin{figure}
\centering
\includegraphics[width=8.0cm]{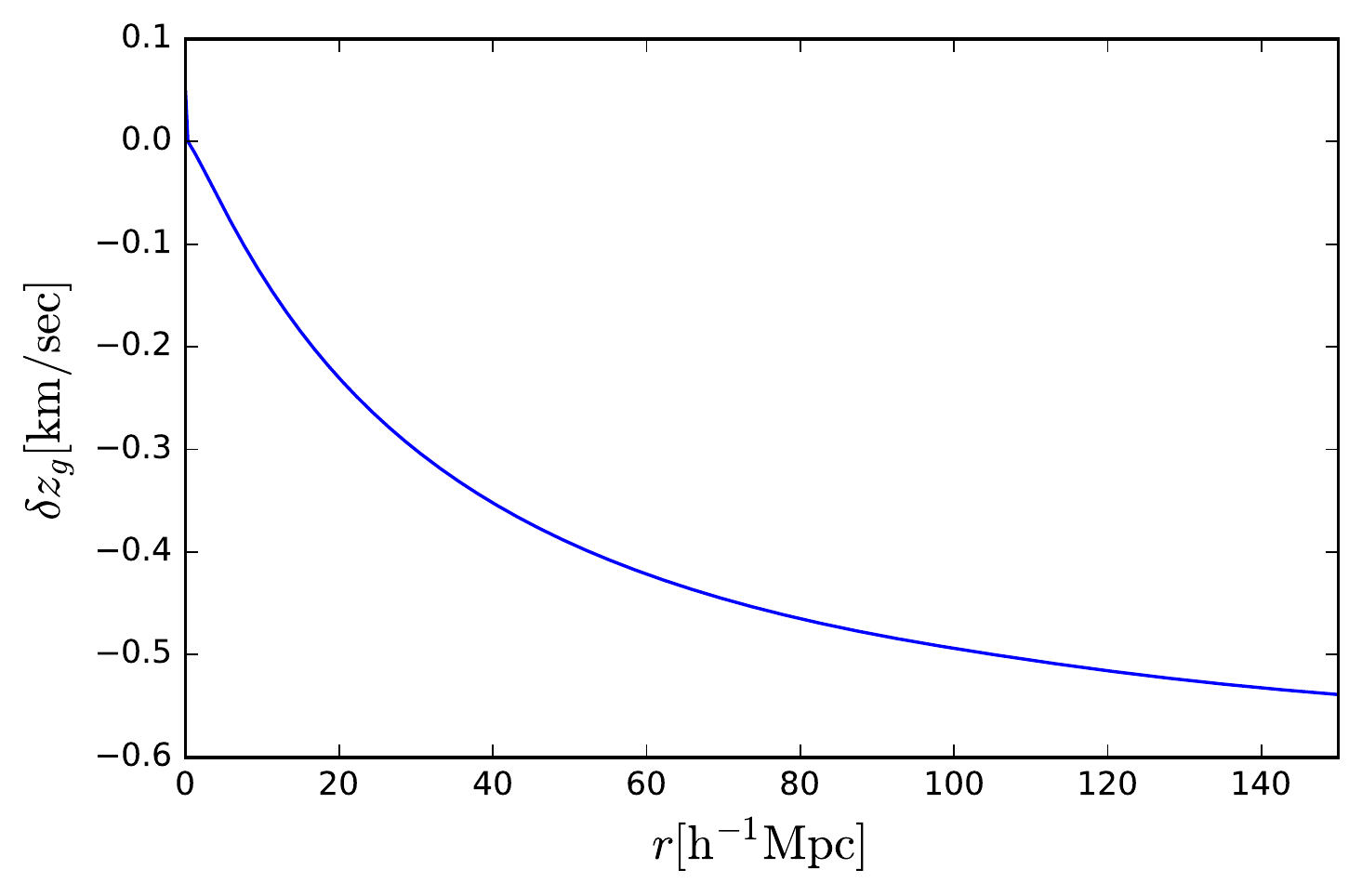}  
\caption{Mean gravitational redshift calculated using eq.~\ref{zg} and considering $b_B=2.25$ and $b_F=2.03$. }
\label{fig:zg}
\end{figure}

Figure \ref{fig:zg} shows the mean gravitational redshift as a function of the separation between $g1$ and $g2$ obtained using eq.~\ref{zg}.

The additional distortion term of the cross-correlation function is due to the peculiar velocities which are caused by two effects: large-scale coherent flows, due to the gravitational instability, 
and small scales random velocity of each galaxy within the cluster. \\
We model the distortion of large-scale cross-correlation function, $\xi_{g1g2}(r_\bot,r_\|)$, where $r_\bot$ and $r_\|$ are the $g1$-$g2$ pairs separation along and across the line-of-sight, as 
described by~\cite{Kaiser01071987} and~\cite{Hamilton:1997zq}:
\begin{figure*}
\begin{tabular}{c c c c}
\includegraphics[width=8.5cm]{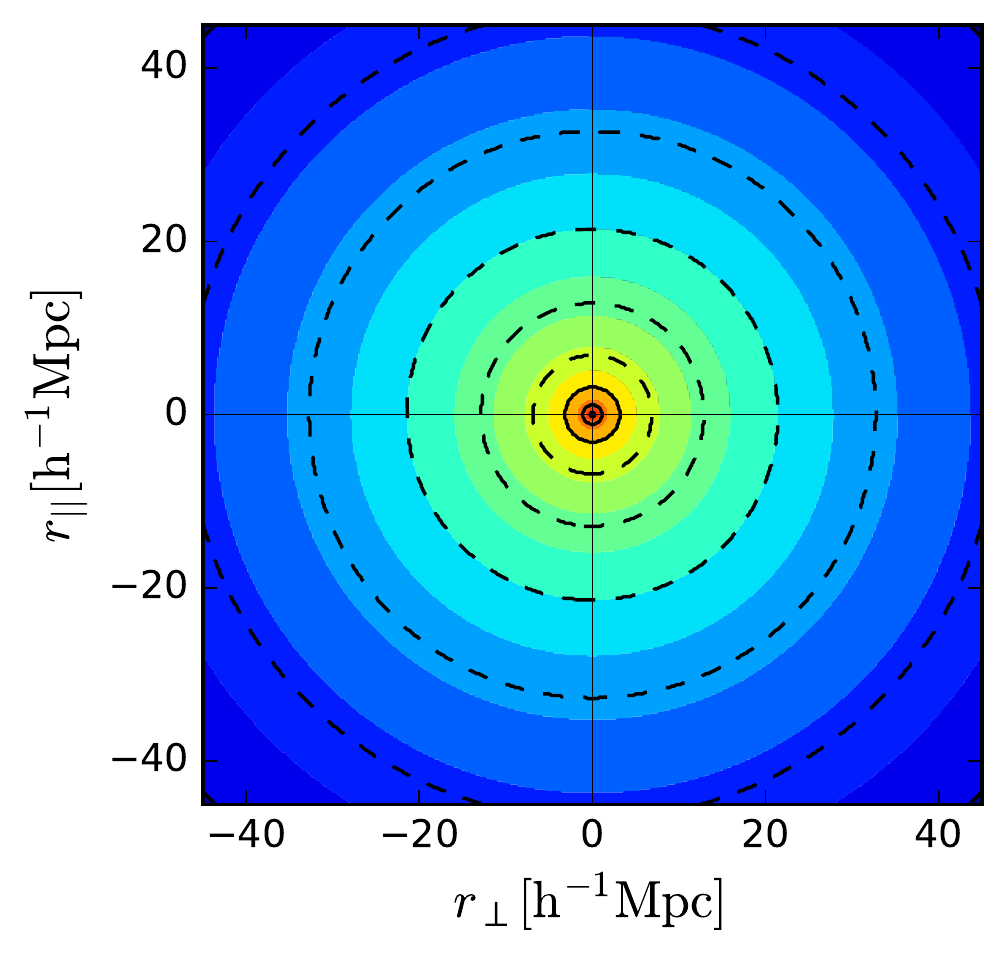}&\includegraphics[width=8.5cm]{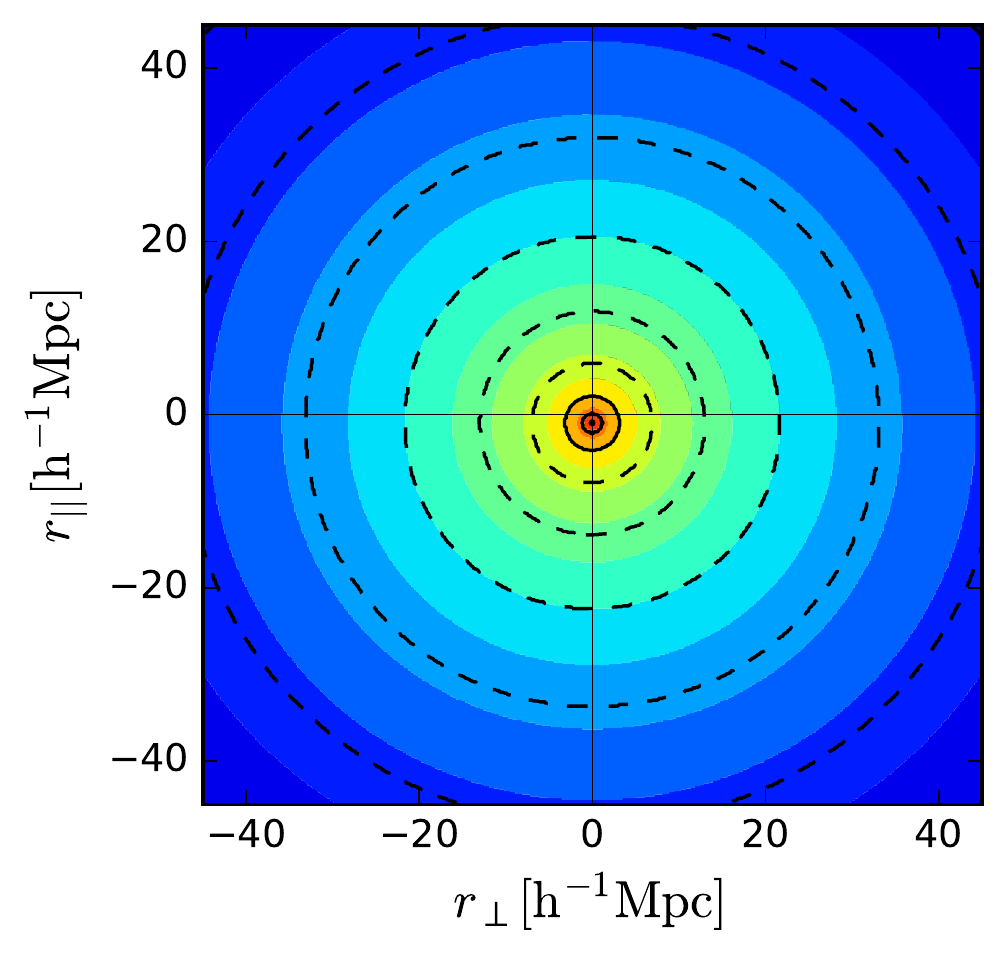}\\
\includegraphics[width=8.5cm]{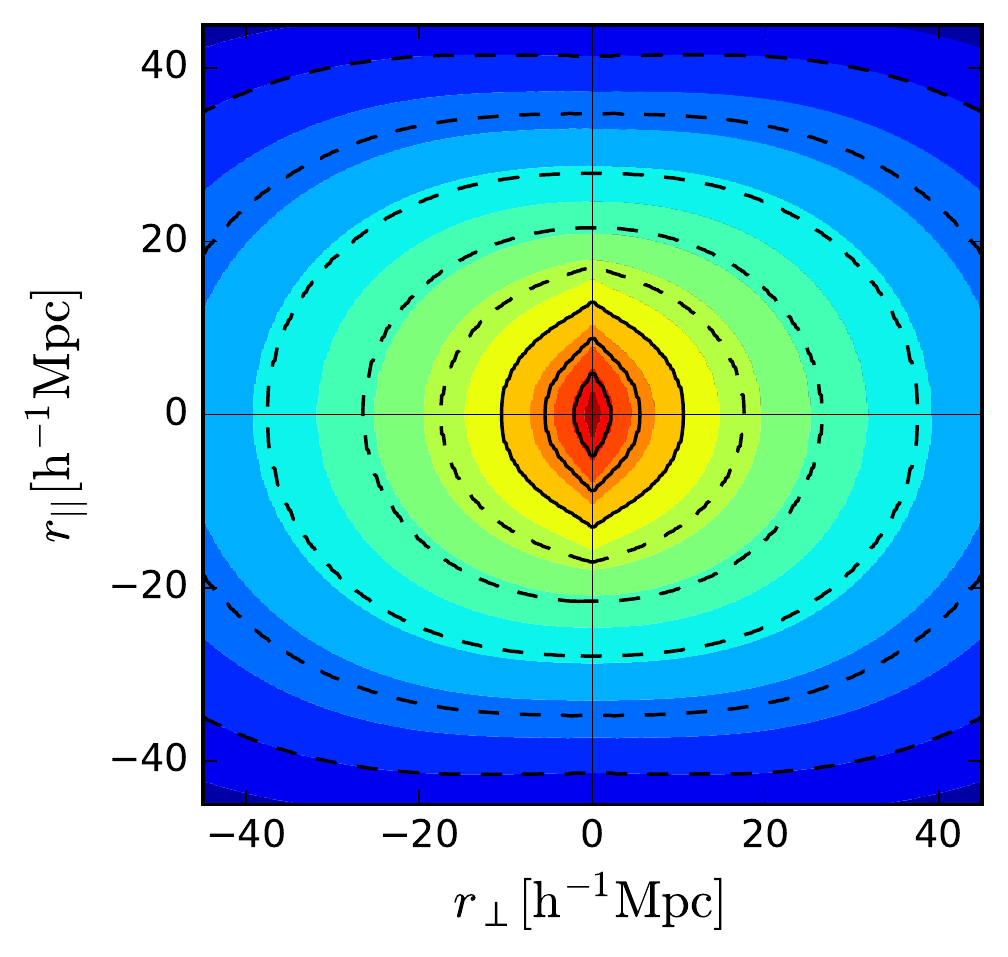}&\includegraphics[width=8.5cm]{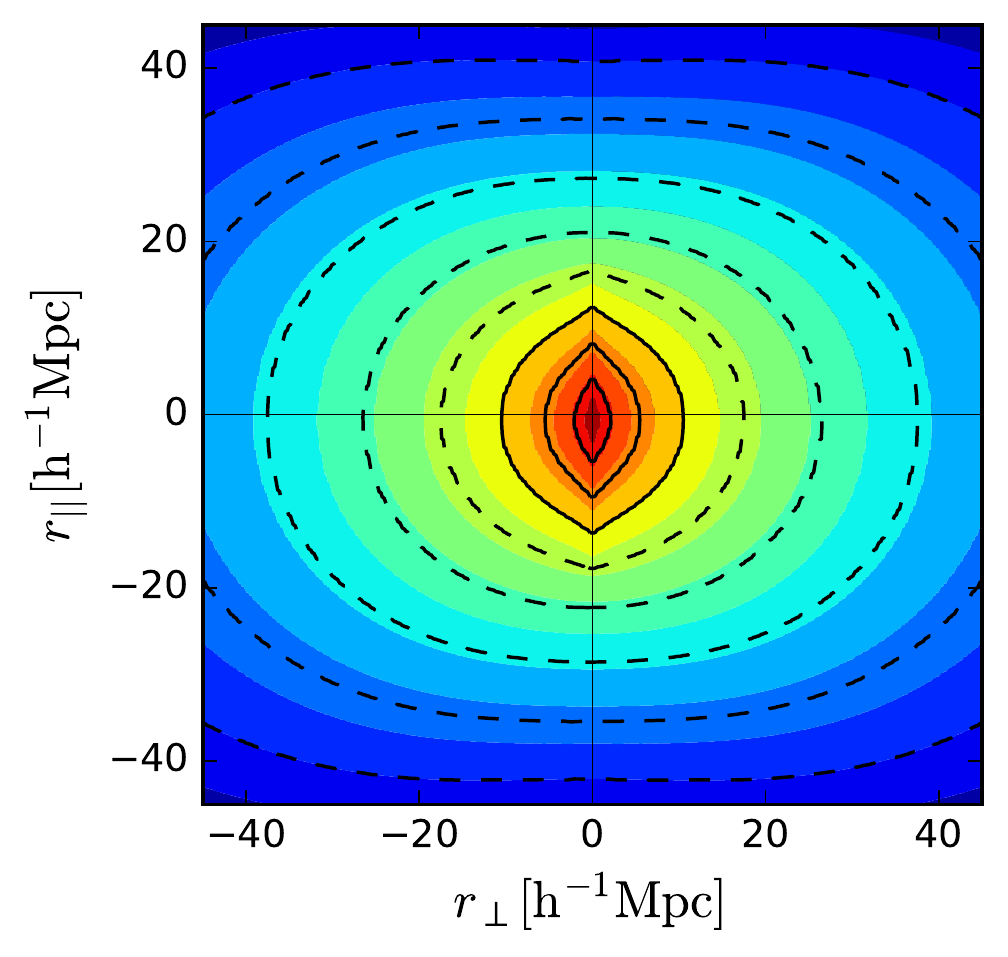}\\
\end{tabular}
 \caption{ Effect of different redshift distortions on cross-correlation function of the two populations of galaxies, $g1$- $g2$, calculated using Newtonian perturbation theory. The top-left panel shows the isotropic correlation function, the top-right panel represents the effect of gravitational redshift, the bottom-left panel illustrates the effect of peculiar velocities and the bottom-right panel shows the two effects combined.}
\label{fig:distortions}
\end{figure*}
\begin{equation}
\xi'_{g1g2}(r_\bot,r_\|)=b_Bb_F[\xi_0(s)P_0(\mu)+\xi_2(s)P_2(\mu)+\xi_4(s)P_4(\mu)]~.
\end{equation}
In the previous equation: $\mu=\cos\theta$, with $\theta$  angle between the pair separation $r$ and the 
line-of-sight, while $P_\ell(\mu)$ are the Legendre Polynomials [$P_0=1$, $P_2=(3\mu^2-1)/2$ and $P_4=(35\mu^4-30\mu^2+3)/8$]. The relations between $\xi_\ell(s)$
and $\xi(r)$ are given by:
\begin{align}
\xi_0(s) &= \left[1 + \frac{1}{3}(\beta_{B}+\beta_{F}) + \frac{1}{5}\beta_{B}\beta_{F}\right] \xi(r)~,&\\
\label{xi0}                                   \nonumber \\
\xi_2(s) &= \left[\frac{2}{3}(\beta_{B}+\beta_{F}) + 
\frac{4}{7}\beta_{B}\beta_{F}\right][\xi(r)-\overline{\xi}(r)]~,&\\         
 \nonumber \\                                   
\xi_4(s) &= \frac{8}{35}\beta_{B}\beta_{F}\left[\xi(r) + \frac{5}{2}\overline{\xi}(r)  
   -\frac{7}{2}\overline{\overline{\xi}}(r)\right]~,&          
\end{align}
with
\be
\overline{\xi}(r) = \frac{3}{r^3}\int^r_0\xi(r')r'{^2}dr',
\nonumber
\ee
\be
\overline{\overline{\xi}}(r) = \frac{5}{r^5}\int^r_0\xi(r')r'{^4}dr'~.
\nonumber
\ee
Here $\beta_B=\Omega^{0.55}/b_B$ and  $\beta_F=\Omega^{0.55}/b_F$ are the redshift space distortion factors that include the large-scale coherent infall. \\
We then convolve our model $\xi'_{g1g2}(r_\bot,r_\|)$ with the pairwise distribution of random velocities, $f(v)$, 
in order to obtain the redshift-space cross-correlation function:
\begin{equation}
\xi_{g1g2}(r_\bot,r_\|)= \int_\infty^\infty f(v)dv\xi'_{g1g2}\left(r_\bot, r_\|-\frac{cz_g(r)-(1+z)v}{H(z)}\right)~.\label{2d}
\end{equation}
We assume that the random peculiar velocity distribution has an exponential form: 
\begin{equation}
f(v)=\frac{1}{\sigma_{12}\sqrt{2}}\exp\left(-\frac{\sqrt{2}|v|}{\sigma_{12}}\right)~,
\end{equation}
where $\sigma_{12}$ is pairwise peculiar velocity dispersion  of g1-g2 galaxies, which we assume to be independent on pair separation. 
Based on simulation results, we choose to set $\sigma_{12}=300$ km/s. 

\begin{figure}
\includegraphics[width=8cm]{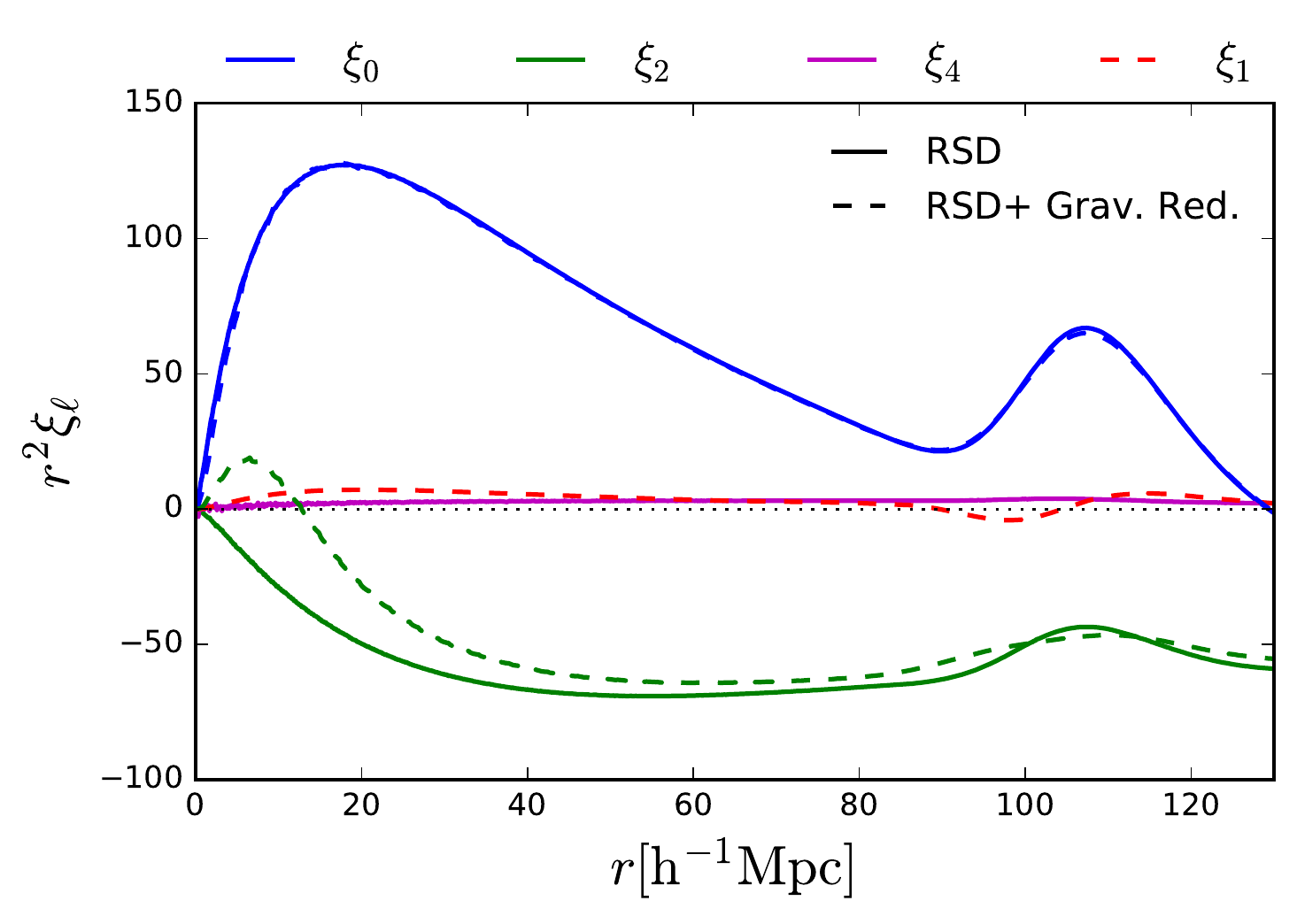}  
\caption{Amplitude of the monopole (blue lines), quadrupole (green lines), hexadecapole (magenta line) and dipole (red line) as a function of the separation, $r$. The solid lines correspond to RSD effect only, while the dashed lines refer to the combination of redshift space distortion and gravitational redshift effect. The dipole is multiplied by a factor of 300 to make it visible. Notice that all the multipoles show a second peak due to the BAO signal. }
\label{fig:decomp} 
\end{figure}

For our model we consider the two populations of galaxies from the previous section with $b_B=2.25$ and $b_F=2.03$
for $g1$ and $g2$ galaxies respectively. We first use eq.~\ref{zg} to distort the cross-correlation function of $g1$ and $g2$ galaxies without considering the effect of redshift space distortions. We then apply only the peculiar velocity term and finally we combine the two terms to study the distortions in $\xi_{g1g2}(r_\bot,r_\|)$.
Figure~\ref{fig:distortions} shows the redshift distortions of the cross-correlation function, $\xi(r_\bot,r_\|)$ of the $g1$ and $g2$ 
galaxies as a function of pair separation along and across the line-of-sight. 

We note that in the absence of distortions (top-left panel), the contours appear circular due to the isotropy of the cross-correlation function. The top-right panel shows the effect of gravitational redshifts multiplied by a factor of 300 for illustrative purpose since this effect is extremely small. We can see that the gravitational redshift leads to a downward displacement of the cross-correlation function contours introducing an asymmetry about the $r_\bot$ axis. 
In the bottom-left panel we consider only the peculiar velocity effect and we can see a stretching of the contours on small scales in the direction of $r_\|$  (the Finger of God effect) and a squashing  of the contours on large scales due to the linear infall (the Kaiser effect). Note that, unlike the gravitational redshift effect, peculiar velocities distortions preserve the symmetry along the line-of-sight direction. Finally the bottom-right panel shows the effect of the inclusion of both the distortion terms. We can observe a flattening of the contour due to the redshift space distortion effect and an increase of asymmetry with respect the $r_\bot$ axis due to the gravitational redshift effect.

\subsection{Extracting multipoles}

We now estimate multipoles of order $\ell$ by decomposing the cross correlation function, $\xi_{g1g2}(r_\bot,r_\|)$, into different modes using Legendre polynomials and integrating over all values of $\mu$:
\be
\xi_\ell(r)=\frac{2\ell+1}{2}\int_{-1}^1\xi_{g1g2}(r_\bot,r_\|) P_\ell(\cos\theta)d\cos\theta~.
\ee
In  this equation $P_\ell(\cos\theta)$ are the Legendre polynomials with $\ell=0,1,2$. In Figure~\ref{fig:decomp} we plot the amplitude of different multipoles as a function of the pair separation $r$ calculated at redshift $z=0.57$ and for $b_B=2.25$ and $b_F=2.03$ computed in~\cite{Alam2016Measurement}. The solid lines show the redshift space distortion effect and the dashed lines illustrate the combination of redshift space distortion and gravitational redshift effects. Moreover the blue lines refer to the monopole, the green lines to the quadrupole, the magenta lines to the hexadecapole and the red line to the dipole. If there are no distortions, the cross-correlation function is isotropic and it is described by only the monopole. Redshift space distortions introduce an anisotropy in the correlation function which is sensitive to the orientation of the galaxies in the pair with respect to the observer. This leads to even order multipoles of correlation function becoming non-zero, most prominently the quadrupole moment. Finally the gravitational redshift breaks the symmetry of the cross-correlation function along the line-of-sight and generates a dipole. In Figure~\ref{fig:decomp} we multiply the dipole by a factor of 300 to better see this effect in comparison to the other ones. As we can see from this figure, the amplitude of the dipole is very small compared to the other multipoles and this means that a measurement of this asymmetry in the cross-correlation function could be difficult on large-scale structure. Note that, unlike the dipole calculated using GRPT, the dipole computed using Newtonian perturbation theory is positive on small scales, $r<50$ Mpc/h, and tends to zero on large scales. 

\begin{figure}
\includegraphics[width=8.5cm]{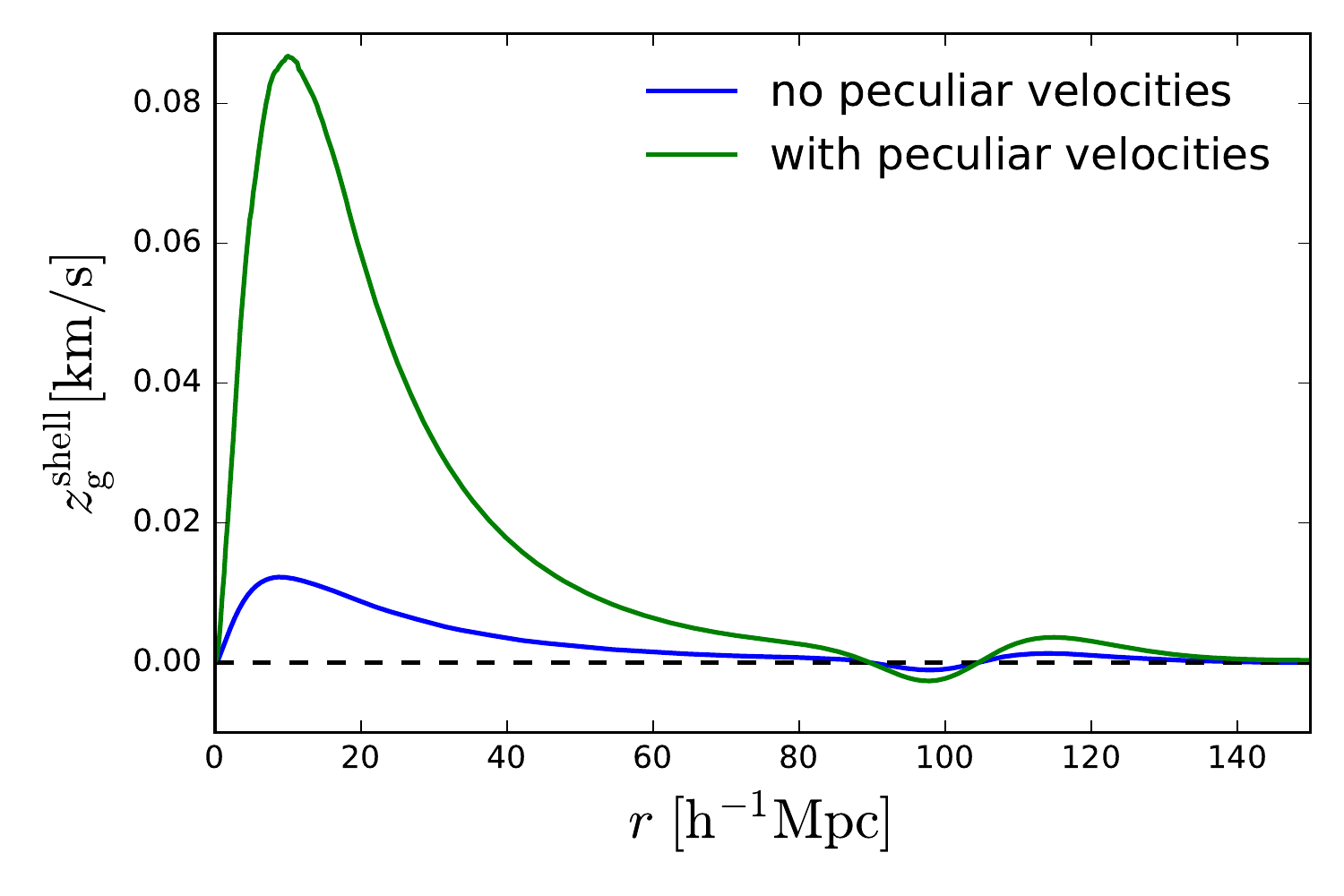}  
\caption{The shell estimator $\zshell$ (eq.~\ref{zshell3}) as a function of the separation, $r$, obtained using Newtonian perturbation theory with (green line) and without (blue line) considering peculiar velocities. }
\label{fig:zshell1}
\end{figure}

\begin{figure}
\centering
\includegraphics[width=9cm]{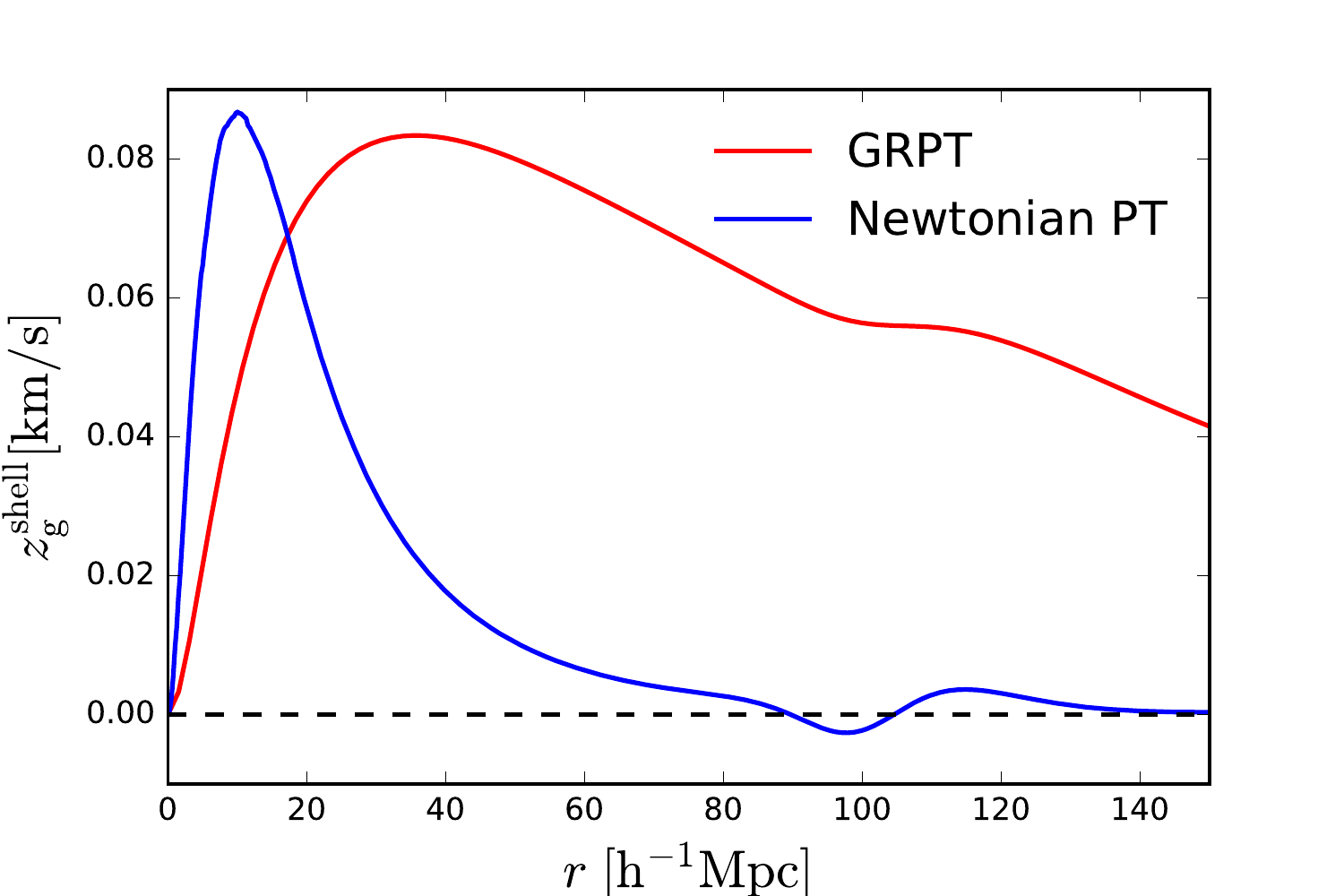}  
\caption{The shell estimator computed from eq.~\ref{zshell3} using General Relativity (red line) and Newtonian perturbation (blue line) theory approach.}.
\label{fig:comp}
\end{figure}

\subsection{The shell estimator}

We use equation~\ref{zshell3} to compute the asymmetry induced by the gravitational redshift. Figure~\ref{fig:zshell1} shows the shell estimator as a function of the pair separation without and with the inclusion of peculiar velocities. We can see that the results obtained considering only the gravitational redshift effect are smaller than that ones obtained including also peculiar velocities. However two curves present a same trend, i.e. it is positive on small scales ($r<20$ Mpc/h) and tends to zero on large linear scales. Figure~\ref{fig:comp} shows the comparison of $\zshell$ using General Relativistic and Newtonian perturbation theory approaches and considering a bias difference of 0.22 as computed in~\cite{Alam2016Measurement}. Notice that the main different between the two lines of Figure~\ref{fig:comp} is that the blue curve includes only the gravitational redshift and redshift space distortion effects, while the red line includes all the relativistic effects that distort the cross-correlation function, such as light cone and Doppler effects. We can also see that Newtonian perturbation theory (blue line) describes well the gravitational redshift on small non-linear scales while GRPT (red line) is able to predict the relativistic effects on large linear scales. 

\begin{figure}
\centering
\includegraphics[width=8.5cm]{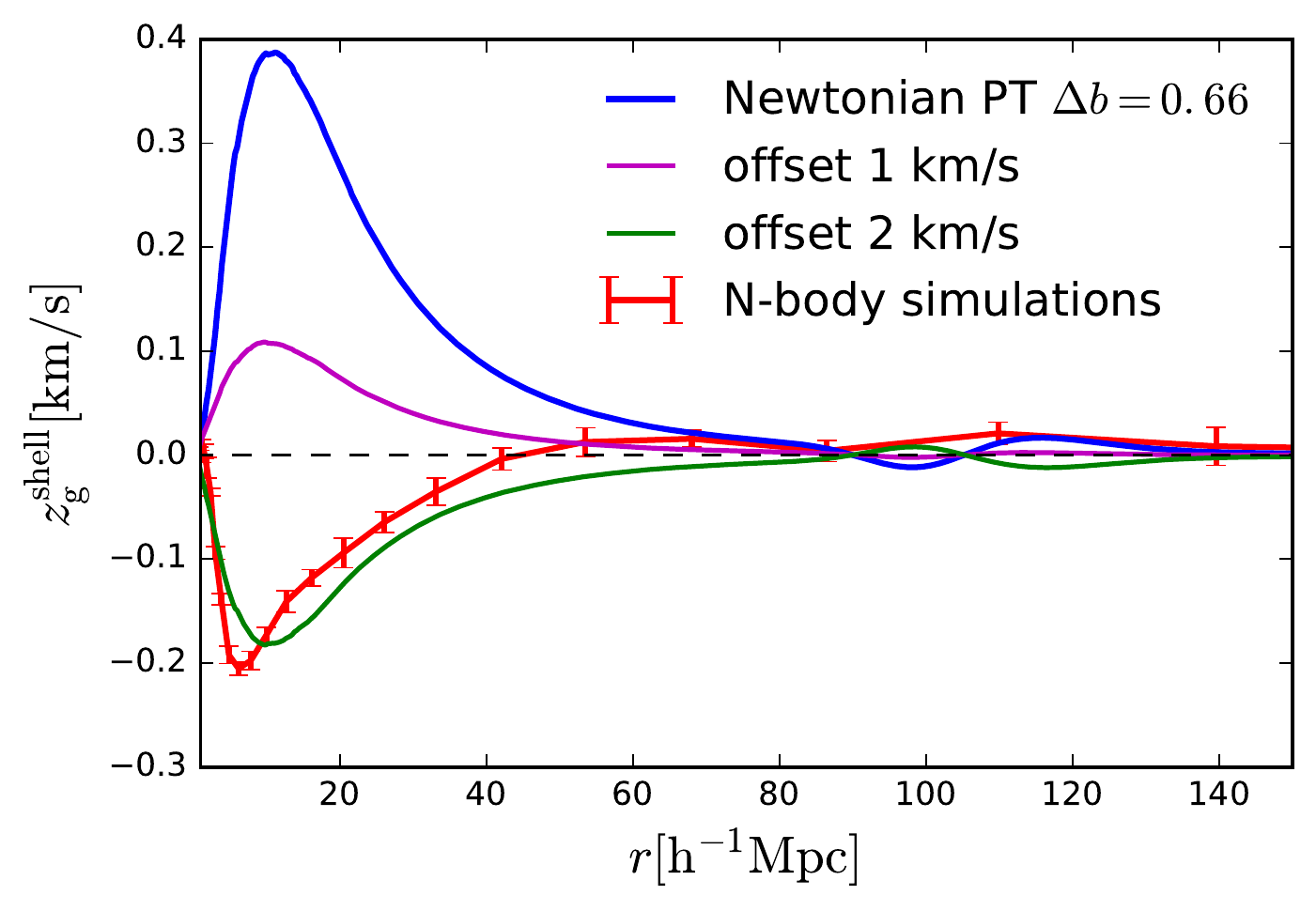}  
\caption{ Comparison between the shell estimator computed using Newtonian perturbation theory (red line) and quasi-Newtonian approach (blue line) and considering a bias difference of 0.66. We also show the results (the magenta and grey lines) obtained after subtracting an offset of 1 and 2 km/s to the $z_g$ component of the redshift in order to mimic the gravitational redshift contribution from the galaxy. }
\label{fig:offset2}
\end{figure}

In Figure~\ref{fig:offset2} we also compare our results for the shell
estimator obtained using Newtonian perturbation theory (blue line),
with those obtained in~\cite{Zhu2016Nbody} using a quasi-Newtonian
approach with N-body simulations (red line). In both  cases we
consider a bias difference of the two galaxy populations of 0.66 as
computed by~\cite{Zhu2016Nbody}. Notice that the sign of~$\zshell$ is
opposite on small scales. In particular it is negative when we use the
quasi-Newtonian simulation method, showing a relative blueshift for galaxy
pairs, and it is positive when we use Newtonian perturbation theory
approach. On the other side, on large scales the shell estimator tends
to zero in both of the approaches. The different behaviour on small
scales is due to the fact that in N-body simulation we include
the information of the structure of the potential well on galactic and
halo scales and we consider the natural scale-dependence of the biases of the two
galaxy sub-samples. On the other hand, in Newtonian perturbation theory approach we assume linear clustering and we model the biases as scale-independent quantities.

In order to approximately mimic the gravitational
redshift contribution from the galaxy, we subtract an offset of 1 and
2 km/s to the $z_g$ component of the redshift (eq.~$\ref{zg}$) before
using it to compute the shell estimator, $\zshell$. By doing this,
we are effectively only modifying the gravitational potential gradient
 at $r=0$.
Nevertheless, this offset has effects on much larger scales as can 
be seen from the Figure. We can see from
Figure~\ref{fig:offset2} (green and magenta lines) that, using this
procedure, we recover a negative sign for the estimator on scales
from $r=0$ Mpc/h all the way to $r\sim 40 $ Mpc/h.
This is agreement
with the~\cite{Zhu2016Nbody} results.


\section{Discussion}
\label{sec:discussion}
In the past few years, relativistic distortions in large-scale structure
have drawn a lot of interest from cosmologists. These effects, together with
the wide-angle and evolution effects, distort the observed
distribution of galaxies inducing line-of-sight asymmetries in the
cross-correlation function of two galaxy populations.
All of the effects mentioned have been studied in various
works~\citep{McDonald:2009ud, yoo2009, yoo2012, Yoo:2014sfa,
  Croft2013, Bonvin:2014owa, Bonvin2014b} and have been observed for
the first time in galaxy clusters by~\cite{Wojtak2011}. The
distortions can be quantified in at least two ways: using the dipole moment of
the cross-correlation function and using the shell
estimator.~\cite{Gaztanaga:2015} have measured both the dipole and shell
estimators from the cross-correlation function of bright and faint
galaxies in the LOWz and CMASS samples of the BOSS DR10 survey and
have shown that the relativistic distortions, which include the
gravitational redshift effect, are not detectable in this survey of
galaxies, at least on scales $r=20 $ Mpc/h and greater. Focussing on smaller
scales (but still in the large-scale structure regime)
~\cite{Alam2016Measurement} have quantified 
the different relativistic effects relevant for two populations of galaxies and
used the BOSS DR12 CMASS galaxy sample to measure these
asymmetries. In particular, they detect the amplitude of relativistic
asymmetry at the significance level of 1.9$\sigma$, 2.5$\sigma$ and
1.7$\sigma$ in the $r$, $i$ and $z$ bands respectively using the shell
estimator and of 2.3$\sigma$ ,0.9$\sigma$, 2.7$\sigma$, 2.8$\sigma$
and 1.9$\sigma$ in the $u$, $g$, $r$, $i$ and $z$ bands respectively,
using the dipole moment. These measurements have peak significance on scales
 around 10 Mpc/h.~\cite{Zhu2016Nbody} have used a
quasi-Newtonian approach, with N-body simulations, to predict these
 asymmetries in the cross-correlation function of two galaxy
different populations and found that the dominant contribution in
the shell estimator is due to gravitational redshift effect.

In this paper, we have analyzed two different approaches in order to
study the effects which induce line-of-sight asymmetries in the
cross-correlation function of two populations of galaxies, examining both
small and large scales ($0<r<150$ Mpc/h). We are particularly
interested in the relativistic contributions which include 
 gravitational redshift. We have computed the two-point
cross-correlation function using GRPT and Newtonian perturbation theory. Following~\cite{Bonvin2014b}
and~\cite{Gaztanaga:2015}, in Sec.~\ref{sec:theory} we have computed
the dipole for the relativistic and wide-angle effects obtained using
the General Relativistic approach and we have also defined the shell
estimator to quantify the line-of-sight anisotropy in the
cross-correlation function in velocity units. In
Sec.~\ref{sec:computing} we have computed the dipole and shell
estimator for a bright and faint population of galaxies using the
effective redshift of BOSS DR12 CMASS sample of galaxies and assuming
different values for the biases. We noticed that the sign of the two
contributions depends on the bias difference, in particular a
positive difference leads the relativistic effect to be dominant over
the wide-angle on all scales, see Figures~\ref{fig:dipole1}
and~\ref{fig:zshell}. Moreover, as already pointed out
by~\cite{Gaztanaga:2015}, we have noted that GRPT
 is able to predict the two effects on very large linear
scales ($r>50$ Mpc/h), but on small scales the same effects, including
the gravitational redshift, are canceled out, showing that this
approach appears to be inconsistent with the observational
results when considering non-linear scales.  In
Sec.~\ref{sec:distortion} we have studied the distortions of the
galaxy cross-correlation function using the Newtonian Perturbation
theory approach. In particular we have explored the distortion in the
cross-correlation function of two galaxy populations induced by
gravitational redshift and peculiar velocities, both separately and
then combined. Figure~\ref{fig:distortions} shows the results in
redshift-space, we noticed that the effect of gravitational redshift
is to shift the contours downwards, corresponding to a relative
blueshift for the g2 (low-mass) galaxies with respect to the g1
(high-mass) ones, while the effect of peculiar velocities is a
squashing of the contours on large scale (Kaiser effect) and a
small scale elongation due to the random velocities (Finger of
God). In order to quantify the line-of-sight asymmetry, we have
computed the dipole moment and the shell estimator of the 2d
cross-correlation function. In both cases we have found a similar
level of asymmetry on small and large scales. In particular we have
noted a positive signal on small scales and a trend consistent with
zero on large scales (in agreements with the results
of~\cite{Gaztanaga:2015} and~\cite{Alam2016Measurement}). By comparing
GRPT with Newtonian perturbation
theory (see Figure~\ref{fig:comp}), we have also found that the first
approach is able to make predictions for relativistic effects on
large, linear scales, while the second one describes successfully the
asymmetry on small, non-linear scales. We have inferred that, on small
scales, the effect of non-linearities, such as the structure of 
potential wells on galactic scales, cannot be treated properly with the
GRPT approach. Finally, we have compared our results for
the shell estimator using Newtonian perturbation theory with those
obtained using a quasi-Newtonian method, with N-body simulations (Figure~\ref{fig:offset2}). We
have noticed that, in the first case, the sign of $\zshell$ on small
scales is positive while in the second case  it is negative. This occurs
because $\zshell$ is sensitive to the gravitational potential of
galaxy on small scales and this is included in N-body simulation
analysis~\citep{Zhu2016Nbody}. Another important difference between
the two analyses is that, with N-body simulations, we included the  
natural
scale-dependence of the bias of the two galaxy sub-samples while in
this paper the biases have been modeled as a scale-independent
quantity. In order to qualitatively 
mimic the gravitational redshift contribution
from the galaxy as part of the Newtonian perturbation theory
approach, we have imposed an offset on the mean gravitational redshift
from eq.~\ref{zg} and we have used it to compute the shell
estimator. With this procedure we have found that the predictions from
N-body simulations and Newtonian perturbation theory have now the same
sign (the shell estimator is positive in both the two
approaches). This shows that the shell estimator is sensitive to
small scales. As any measurement
of the clustering asymmetry relies on relative redshift measurements 
of one galaxy from another, any redshift difference on any scale  ( including
on galactic scales) can have an additive effect. This means that 
calculational approaches, such as perturbation theory which rely on 
large scale averaging and do not include the small scale structure
in the density and velocity field, may not converge as rapidly as hoped
to the correct result.

\begin{figure}
\centering
\includegraphics[width=8.5cm]{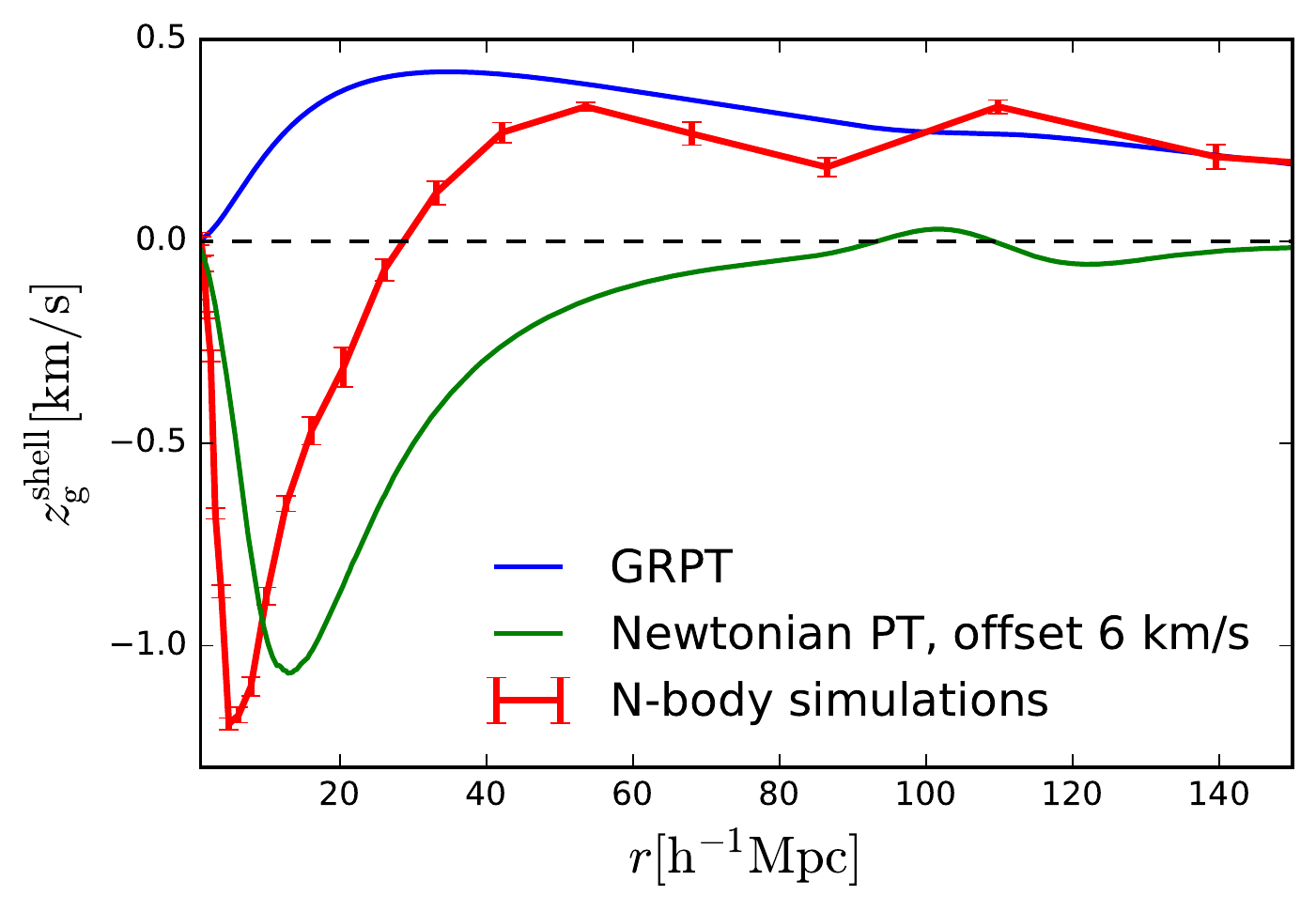}  
\caption{ Comparison between the shell estimator computed using linear perturbation theory (blue line), Newtonian perturbation theory (green line) and quasi-Newtonian approach (red line) and considering a bias difference of 0.66.}
\label{fig:comparison}
\end{figure}

It is important  to 
point out that there is 
no inconsistency between the ``Newtonian'' techniques and GRPT
approaches on large scales.
In the present paper, if we had added additional terms to our
Newtonian calculation to model the luminosity distance perturbation (rather
than just including gravitational redshift), then this would 
have resolved the mismatch between the Newtonian theory and GRPT on large scales seen in Figure~\ref{fig:comp}. 
  This is illustrated by  
  Figure~\ref{fig:comparison}, where we compare the results obtained
  using Newtonian perturbation theory (green line) and GRPT (blue
  line), with those obtained by~\cite{Zhu2016Nbody} using N-body
  simulations (red line). The latter includes all relativistic effects
  (gravitational redshift, special relativistic beaming, light cone,
  transverse Doppler and luminosity distance perturbation), with 
luminosity distance perturbation being dominant on large scales.
  We can see that on large scales ($r > 40$ Mpc/h), the shell estimator
  computed from simulations is in agreement with that computed using
  linear perturbation theory. However the Newtonian perturbation
  theory approach we studied in this work, only includes the
  distortion induced by gravitational redshifts and peculiar
  velocities. A complete analytic treatment of all relativistic
effects in the Newtonian framework is beyond the scope of this
  paper.

We conclude that a combination of General Relativistic perturbation
theory and non-linear techniques is needed to study the asymmetry due
to relativistic effects on large and small scales. As shown
in~\cite{Zhu2016Nbody}, there also exist important uncertainties in
the theoretical predictions, such as the precise nature of structure
on galactic scales, that is necessary to model in order to explore
these effects with accuracy. Finally, we recognize the increasing
role which
that measurements of relativistic effects will have in cosmology.
Gravitational redshifts  in particular are measurable with
current data and will be useful for
constraining models, such as modified theories of gravity.

\section*{Acknowledgments}
E.G. kindly acknowledges S\'ebastien Fromenteau and Simone Ferraro for interesting
discussions. E.G. also thanks Mauricio Reyes Hurtado and Giuseppe Iacobellis for useful comments. This work was supported by NSF grant AST1412966. SA and SH are supported by NASA grants 12-EUCLID11-0004. SA is also supported by the European Research Council through the COSFORM Research Grant ($\#$670193).

\bibliography{Master_Elena}
\bibliographystyle{mnras}

\label{lastpage}

\end{document}